\documentclass[
twocolumn,
pra,showpacs,preprintnumbers,amsmath,amssymb]{revtex4}

\usepackage{graphicx}
\usepackage{amssymb}
\usepackage{amsmath}
\usepackage{dsfont}
\usepackage{bm}
\usepackage{mathrsfs}

\begin{document}

\title{Energy-momentum balance in quantum dielectrics}
\author{Ulf Leonhardt}
\affiliation{School of Physics and Astronomy, University of St Andrews, 
North Haugh, St Andrews, Fife, KY16 9SS, Scotland}
\begin{abstract}
We calculate the energy-momentum balance in quantum 
dielectrics such as Bose-Einstein condensates.
In agreement with the experiment 
[G. K. Campbell {\it et al.}
Phys. Rev. Lett. {\bf 94}, 170403 (2005)]
variations of the Minkowski momentum
are imprinted onto the phase, 
whereas the Abraham tensor drives the flow of the dielectric.
Our analysis indicates that
the Abraham-Minkowski controversy
has its root in the R\"ontgen interaction 
of the electromagnetic field in dielectric media. 
\end{abstract}
\date{\today}
\maketitle


\section{Introduction}

It is surprising \cite{Peierls}
that the momentum of light in media has been subject to considerable 
debate \cite{Peierls,Minkowski,Abraham,Debate}
for almost a century.
Moreover, although the main contenders,
the Minkowski momentum \cite{Minkowski}
${\bf D}\times{\bf B}$
and the Abraham momentum \cite{Abraham}
${\bf E}\times{\bf H}/c^2$,
differ by a substantial factor,
the refractive index squared,
precise experimental tests of this thorny issue
have been scarce \cite{Experiments,Campbell}.
Only recently, the photon recoil momentum 
was directly measured \cite{Campbell}
in a medium made of a Bose-Einstein condensate
\cite{BEC}.
From a theoretical point of view,
a condensate is a very simple system,
a nearly ideal quantum gas or irrotational fluid \cite{BEC}
that allows studies of fundamental effects
without many complications from material details.
From a practical perspective,
the momentum transfer of light in condensates 
is important in high-precision atom interferometry
\cite{Campbell}.

Here we deduce the energy-momentum balance
in such quantum dielectrics from geometric principles
that do not depend much on microscopic details.
We find that,
in the non-relativistic limit,
for a condensate with number density $\varrho$, phase $S$,
electric permittivity $\varepsilon$
and magnetic permeability $\mu$ the 
total momentum density is
\begin{equation}
\label{eq:g1}
{\bf g} = \varrho\,\hbar\nabla S + {\bf D}\times{\bf B}
\,.
\end{equation}
This result shows that the Minkowski momentum 
is imprinted into the phase of the quantum dielectric,
in agreement with the experiment \cite{Campbell}.
However, we can also express the total momentum density
in the Abraham form
\begin{equation}
\label{eq:g2}
{\bf g}= {\varrho}\,{m}{\bf u} +
 \frac{{\bf E}\times{\bf H}}{c^2}
\end{equation}
where $m$ is the atomic mass,
defining the quantity 
\begin{equation}
\label{eq:u}
{\bf u} = \frac{1}{m}\left[
\hbar\nabla S + \left(\varepsilon-\frac{1}{\mu}\right)
\frac{{\bf E}\times{\bf B}}{\varrho}\right]
\end{equation}
and using the constitutive equations of the electromagnetic field
in a medium at rest,
\begin{equation}
\label{eq:const0}
{\bf D} = \varepsilon_0\varepsilon {\bf E} \,,\quad
{\bf H} = \frac{\varepsilon_0 c^2}{\mu} {\bf B}
\,.
\end{equation}
The important point of this simple exercise in re-expressing
${\bf g}$ is the physical meaning of ${\bf u}$:
for dielectric media, 
${\bf u}$ describes the flow velocity.
The mechanical momentum $m{\bf u}$
differs from the canonical momentum $\hbar\nabla S$ by the
R\"ontgen term
$(\varepsilon-1/\mu){\bf E}\times{\bf B}/\varrho$
that, for neutral atoms,
plays the role of the vector potential
for charged particles \cite{Roentgen}.

The R\"ontgen interaction  \cite{Roentgen} stems from the 
behavior of the moving atomic dipoles 
that constitute the medium: when set in motion
induced electric dipoles perceive mixtures of 
electric and magnetic fields, 
according to the Lorentz transformations 
of the fields \cite{LL8},
and the same applies to magnetic dipoles.
The R\"ontgen interaction  \cite{Roentgen} 
is most easily deduced from
the Lagrangian of a classical induced dipole
\begin{equation}
L = \frac{m}{2}u^2 + \frac{\alpha_E}{2}E'^2
+ \frac{\alpha_B}{2}c^2 B'^2
\end{equation}
where $\alpha_E$ and $\alpha_B$ denote the 
electric and magnetic polarizabilities.
The electromagnetic field interacts with the moving dipole
in the locally co-moving frame, indicated by primes, 
where the dipole is at rest. 
In lowest order, we obtain from the 
Lorentz transformations \cite{LL8},
${\bf E}' = {\bf E} + {\bf u}\times {\bf B}$, 
${\bf B}' = {\bf B} - {\bf u}\times {\bf E}/c^2$
the canonical momentum 
\begin{equation}
{\bf p} = \frac{\partial L}{\partial {\bf u}} =
m{\bf u} -(\alpha_E+\alpha_B)  ({\bf E}\times {\bf B}) \,.
\end{equation}
For a medium with dipole density $\varrho$,
the electric and magnetic polarizabilities
give rise to the electric permittivity 
and magnetic permeability
\begin{equation}
\label{eq:pol}
\varepsilon = 1 + \frac{\alpha_E}{\varepsilon_0}\varrho
\,,\quad
\mu = 1 - \frac{\alpha_B}{\varepsilon_0}\varrho 
\,.
\end{equation}
For a quantum dielectric, the canonical momentum 
corresponds to the phase gradient, 
which gives Eq.\ (\ref{eq:u}).
 
Formulae (\ref{eq:g1}-\ref{eq:u}) illustrate
the ambivalent nature of the electromagnetic momentum
in media: 
variations of the Minkowski momentum
are imprinted onto the phase, 
whereas the Abraham momentum drives the flow
of the quantum dielectric.  
Flow and phase gradient differ by the R\"{o}ntgen term.
Consequently,
the R\"{o}ntgen interaction \cite{Roentgen}
appears to be at the heart of the Abraham-Minkowski 
controversy \cite{Debate}. 

In this paper, we deduce and generalize 
the relationships (\ref{eq:g1}-\ref{eq:u})
in a relativistically covariant form.
For our analysis we borrow ideas 
from General Relativity \cite{LL2}
that have their historic precursors in
Fermat's Principle in geometric optics \cite{BornWolf}:
light rays take the shortest optical paths in a medium,
regardless how curved their trajectories are.
In terms of General Relativity \cite{LL2},
the rays follow geodesic lines with respect
to a metric that is proportional 
to the refractive index.
In general, the electromagnetic field perceives
isotropic, non-dispersive and possibly moving media
as effective space-time geometries \cite{Gordon,Leo}.
On the other hand, in atom optics
the roles of light and matter are often reversed:
light acts as a medium for matter waves.
It is therefore natural to postulate \cite{Leo}
that quantum dielectrics perceive the
electromagnetic field as an effective
space-time geometry as well.
This approach turned out to be completely
consistent with Maxwell's equations \cite{Leo},
which justifies the idea.
However, our previous calculation 
of the energy-momentum tensor
was indirect, flawed \cite{Flaw}
and gave the wrong energy-momentum tensor of the 
dielectric matter.
Here we calculate the energy-momentum balance
directly and show how the Minkowski and 
Abraham forms are related to each other, considering
the dynamics of the medium. 
In Sec. II we summarize the essence of the geometric approach
to quantum dielectrics.
Section III deduces the Minkowski form 
of the energy-momentum balance, 
both proving and generalizing the result
(\ref{eq:g1}) in the non-relativistic limit.
Section IV casts the total energy-momentum tensor
in Abraham form, expressing the general connection
between the two forms.

\section{Geometric approach}

For simplicity, we entirely focus on the dielectric
interaction of the atomic condensate 
with the electromagnetic field.
The collisional contact interaction of the atoms 
and any external potentials can be easily
included as additional terms in the 
energy-momentum balance and hence
they are not considered here.
We use the relativistic notation of Ref.\ \cite{Leo}
with Greek indices referring to the space-time
coordinates $x^\alpha$, but we restrict ourselves
to Cartesian coordinates in flat Minkowski space-time
with metric tensor 
$g_{\alpha\beta}=\mbox{diag}(1,-1,-1,-1)$.
Partial derivatives with respect to the coordinates
$x^\alpha$ are denoted by $\partial_\alpha$.
Throughout this paper we employ Einstein's summation
convention over repeated indices.
The electromagnetic field is described by the antisymmetric
field-strength tensor $F_{\alpha\beta}$ in SI units \cite{Leo}.

We require that the quantum dielectric perceives
the electromagnetic field as an effective space-time geometry
with the atomic metric tensor $g_{\alpha\beta}^A$
that is relativistically covariant.
In locally co-moving Galilean coordinates $g_{\alpha\beta}^A$
contains the dipole potentials of the atoms,
$-(\alpha_E/2)E^2$ and $-(\alpha_B/2)c^2B^2$.
Consequently, $g_{\alpha\beta}^A$ 
must be quadratic in the field strengths. 
This, combined with general covariance, leads to 
the only option \cite{Leo}
\begin{equation}
\label{eq:atommetric}
g^A_{\alpha\beta} = \left( 1 - a\, {\mathscr L}_F \right) 
g_{\alpha\beta} - b\,T^F_{\alpha\beta}
\end{equation}
where ${\mathscr L}_F$ 
denotes the free-field Lagrangian density \cite{LL2}
and $T^F_{\alpha\beta}$ 
the free energy-momentum tensor \cite{LL2},
\begin{eqnarray}
{\mathscr L}_F &=& 
- \frac{\varepsilon_0}{4}\, F_{\alpha\beta} F^{\alpha\beta}
\,,\nonumber\\
T^F_{\alpha\beta} &=&
\varepsilon_0 F_{\alpha\alpha'} g^{\alpha'\beta'} F_{\beta'\beta} -
{\mathscr L}_F g_{\alpha\beta}
\,.
\label{eq:tensor}
\end{eqnarray}
We consider the condensate in the hydrodynamic limit \cite{BEC} 
subject to the standard Lagrangian density
\begin{equation}
\label{eq:al}
{\mathscr L}_A = 
\sqrt{-g_A}\, |\psi|^2
\left(
\frac{\hbar^2}{2m} g_A^{\alpha\beta}
(\partial_\alpha S)(\partial_\beta S) - \frac{mc^2}{2}
\right)
\,.
\end{equation}
Here $m$ denotes the atomic mass,
$g_A$ is the determinant and $g_A^{\alpha\beta}$
the inverse of the metric tensor $g^A_{\alpha\beta}$.
The coefficients $a$ and $b$ in $g^A_{\alpha\beta}$
are obtained from considering the non-relativistic limit
of the atomic Lagrangian density (\ref{eq:al}).
They turn out to be related to the electric and magnetic 
polarizabilities $\alpha_E$ and $\alpha_B$ as \cite{Leo}
\begin{equation}
a = \frac{\alpha_E-\alpha_B}{\varepsilon_0 mc^2}
\,,\quad
b = \frac{\alpha_E+\alpha_B}{\varepsilon_0 mc^2}
\,.
\end{equation}
The Euler-Lagrange equations give
the Hamilton-Jacobi equation
\begin{equation}
\label{eq:hj}
g_A^{\alpha\beta} 
(\partial_\alpha S) 
(\partial_\beta S) =
\frac{m^2c^2}{\hbar^2} 
\end{equation}
and the equation of continuity
\begin{equation}
\label{eq:cont}
\partial_\alpha \sqrt{-g_A}\, |\psi|^2 
g_A^{\alpha\beta} \partial_\beta S = 0
\,.
\end{equation}
Consider the four-vectors
\begin{equation}
\label{eq:wa}
w_\alpha = \frac{\hbar}{mc}\, \partial_\alpha S
\,,\quad
w^{(\alpha)} = g_A^{\alpha\beta} w_{\beta}
\,.
\end{equation}
If $g_A^{\alpha\beta}$ were the true space-time metric
in which the medium moves
$w^{(\alpha)}$ would be the four-velocity
of the dielectric, because it 
enters as a four-velocity in the equation of continuity
(\ref{eq:cont}) and is normalized to unity
with respect to the metric (\ref{eq:atommetric}).
We obtain from 
the Hamilton-Jacobi equation (\ref{eq:hj})
the geodesic equation
\begin{equation}
\label{eq:geo}
w^{(\alpha)}\partial_\alpha w_\nu = 
\frac{1}{2} w^{(\alpha)}w^{(\beta)}
\partial_\nu g^A_{\alpha\beta}
\end{equation}
that plays the role of the Euler equation 
of the dielectric fluid \cite{LL6}.
However, the true four-velocity $u^\alpha$
differs from $w^{(\alpha)}$ by the norm of $w^{(\alpha)}$
with respect to the metric $g_{\alpha\beta}$,
\begin{equation}
\label{eq:udef}
u^\alpha = \frac{w^{(\alpha)}}{w}
\,,\quad
w=\sqrt{g_{\alpha\beta}\,w^{(\alpha)}w^{(\beta)}}
\,.
\end{equation}
In a realistic dielectric, the norm $w$ of the
quasi-four-velocity $w^{(\alpha)}$ is nearly unity,
up to corrections in the order of the 
electromagnetic field energy divided 
by $mc^2$ \cite{Leo}.
Note that such corrections matter in expressions
that contain the total energy including
the rest energy.
The equations of continuity imply that the 
atom-number density $\varrho$ is not $|\psi|^2$, but
\begin{equation}
\label{eq:rho}
\varrho = \sqrt{-g_A}\,w\,|\psi|^2
\,.
\end{equation}
In the non-relativistic limit, Eq.\ (\ref{eq:udef})
reduces to the relationship (\ref{eq:u}) between
the three-dimensional velocity ${\bf u}$ and the 
gradient of the phase $S$ including the 
R\"ontgen term. 
In other words, 
the difference between co- and contravariant
velocity vectors in the effective geometry 
(\ref{eq:atommetric})
accounts for the R\"ontgen interaction
\cite{Roentgen}.

Finally, we note that our geometric approach 
is justified \cite{Leo}, because the
total Lagrangian density 
${\mathscr L}_A+{\mathscr L}_F$
that generates the equations of motion
(\ref{eq:hj}) and (\ref{eq:cont}) 
also generates Maxwell's equations 
\begin{equation}
\label{eq:maxwell}
\partial_\alpha H^{\alpha\beta} = 0
\,.
\end{equation}
Here $H^{\alpha\beta}$ contains the electromagnetic
${\bf D}$ and ${\bf H}$ fields in SI units \cite{Leo}
with the constitutive equations
\begin{equation}
\label{eq:const}
H^{\alpha\beta} = 
\frac{\varepsilon_0}{\mu}\,
g_F^{\alpha\alpha'}
g_F^{\beta\beta'}
F_{\alpha'\beta'}
\end{equation}
expressed in terms of Gordon's metric \cite{Gordon,Leo}
\begin{equation}
\label{eq:gordon}
g_F^{\alpha\alpha'} =
g^{\alpha\alpha'} + (\varepsilon\mu-1)u^\alpha u^\beta
\end{equation}
with the local four-velocity (\ref{eq:udef})
and the electric permittivity $\varepsilon$
and magnetic permeability $\mu$, 
\begin{eqnarray}
\varepsilon &=& 1 + \frac{a+b}{2}\,mc^2 w\varrho =
1 + \frac{\alpha_E}{\varepsilon_0}\,w\varrho \,,
\nonumber\\
\frac{1}{\mu} &=& 1 + \frac{a-b}{2}\,mc^2 w\varrho =
1 - \frac{\alpha_B}{\varepsilon_0}\,w\varrho \,,
\label{eq:ab}
\end{eqnarray}
that are consistent with  the polarizabilities (\ref{eq:pol})
in the realistic case where $w$ approaches unity.
Equations (\ref{eq:const}-\ref{eq:gordon})
correspond to the constitutive equations (\ref{eq:const0})
in locally co-moving frames.
Light and matter perceive each other as effective space-time
geometries where Gordon's metric (\ref{eq:gordon})
characterizes the geometry seen by the electromagnetic field
and the atomic metric (\ref{eq:atommetric})
describes the effective geometry of the quantum dielectric.

\section{Minkowski tensor.}

In order to find the energy-momentum tensor
for the dielectric,
consider the tensor that would represent 
the energy-momentum of the dielectric matter
if the effective metric $g_A^{\alpha\beta}$ describes 
the true space-time geometry,
\begin{equation}
\label{eq:theta}
\Theta_\nu^\alpha = mc^2|\psi|^2 w^{(\alpha)} w_\nu
\,.
\end{equation}
We obtain from the equations of motion
(\ref{eq:cont}) and (\ref{eq:geo})
\begin{eqnarray}
\partial_\alpha\sqrt{-g_A}\,\Theta_\nu^\alpha 
&=&
\sqrt{-g_A}\,\frac{mc^2}{2}\, |\psi|^2\,
w^{(\alpha)} w^{(\beta)} 
\partial_\nu g^A_{\alpha\beta}
\nonumber\\
&=&
\sqrt{-g_A}\,\frac{1}{2}\,\Theta_\beta^\alpha\,
g_A^{\beta\beta'} \partial_\nu g^A_{\alpha\beta'} 
\,,
\label{eq:cl}
\end{eqnarray}
which corresponds to a conservation law in a curved
space-time geometry \cite{LL2} where
the right-hand side accounts for the inertial forces.
In our case, these are the forces that light exerts on
the dielectric medium.

In the following we show that the dielectric forces
stem from the Minkowski energy-momentum tensor
of the electromagnetic field \cite{Minkowski}.
The Minkowski tensor is defined as \cite{Leo}
\begin{equation}
\label{eq:mk}
T_{\mathrm{Mk}\nu}^\alpha =
H^{\alpha\alpha'} F_{\alpha'\nu} + 
\frac{1}{4}\,H^{\alpha'\beta'}F_{\alpha'\beta'}
\delta_\nu^\alpha
\,.
\end{equation}
We obtain from Maxwell's equations (\ref{eq:maxwell})
\begin{equation}
\partial_\alpha T_{\mathrm{Mk}\nu}^\alpha =
H^{\alpha\beta} \partial_\alpha F_{\beta\nu} +
\frac{1}{4}\, \partial_\nu H^{\alpha\beta} F_{\alpha\beta}
\,.
\end{equation}
We apply the Bianchi identity \cite{LL2}
of the field-strength tensor, 
\begin{equation}
\partial_\nu F_{\alpha\beta} + 
\partial_\alpha F_{\beta\nu} +
\partial_\beta F_{\nu\alpha} = 0 \,,
\end{equation}
that implies
\begin{equation}
H^{\alpha\beta}
\left(2\partial_\alpha F_{\beta\nu}+
\partial_\nu F_{\alpha\beta}
\right) =0\,.
\end{equation}
We use Gordon's metric form of the 
constitutive equations (\ref{eq:const}) 
and Eq.\ (\ref{eq:ab}) for $\varepsilon$ and $\mu$
to obtain
\begin{eqnarray}
\partial_\alpha T_{\mathrm{Mk}\nu}^\alpha 
&=&
\frac{1}{4}\,F_{\alpha'\beta'} F_{\alpha\beta}\, \partial_\nu
\frac{\varepsilon_0}{\mu}\,g_F^{\alpha\alpha'} g_F^{\beta\beta'}
\nonumber\\
&=&
\frac{\varepsilon_0}{2}\,
F_{\alpha'\beta'} F_{\alpha\beta}\, \partial_\nu
\left(\varepsilon-\frac{1}{\mu}\right) 
g^{\alpha\alpha'} u^\beta u^{\beta'} 
- {\mathscr L}_F \partial_\nu \frac{1}{\mu}
\nonumber\\
&=&
\frac{\varepsilon_0}{2}\,
mc^2 b\, F_{\alpha'\beta'} F_{\alpha\beta}\,
g^{\alpha\alpha'} \partial_\nu w\varrho u^\beta u^{\beta'}
- {\mathscr L}_F \partial_\nu \frac{1}{\mu}
\nonumber\\
&=&
-\frac{mc^2}{2}\,b
\left(T^F_{\beta\beta'} + {\mathscr L}_F g_{\beta\beta'}\right)
\partial_\nu w\varrho u^\beta u^{\beta'}
\nonumber\\
&& - {\mathscr L}_F \partial_\nu \frac{1}{\mu} \,.
\end{eqnarray}
We use the relation
\begin{equation}
\partial_\nu \frac{1}{\mu}
=
\frac{mc^2}{2}(a-b)\, g_{\beta\beta'}\,\partial_\nu
w\varrho u^\beta u^{\beta'}
\end{equation}
and the definition (\ref{eq:atommetric}) of 
$g^A_{\alpha\beta}$ to 
arrive at the expressions
\begin{eqnarray}
\partial_\alpha T_{\mathrm{Mk}\nu}^\alpha 
&=&
\frac{mc^2}{2}\left(g^A_{\alpha\beta}-g_{\alpha\beta}\right)
\partial_\nu w\varrho u^\alpha u^\beta
\nonumber\\
&=&
-\frac{mc^2}{2}\, w\varrho u^\alpha u^\beta 
\partial_\nu g^A_{\alpha\beta} + \partial_\nu p
\,.
\end{eqnarray}
Here $p$ denotes the dielectric pressure \cite{Leo}
\begin{equation}
\label{eq:p}
p =\frac{mc^2}{2}\,
\left(g^A_{\alpha\beta}-g_{\alpha\beta}\right)
w\varrho u^\alpha u^\beta 
\,.
\end{equation}
Using the definition (\ref{eq:atommetric}) 
of $g^A_{\alpha\beta}$ 
and Eqs.\ (\ref{eq:const}-\ref{eq:ab})
one can express the pressure $p$ 
in terms of the electromagnetic fields  \cite{Leo}
\begin{equation}
\label{eq:pp}
p =
\frac{1}{4}F_{\alpha\beta}
\left(H^{\alpha\beta}-\varepsilon_0 F^{\alpha\beta}\right)
\end{equation}
where in the non-relativistic limit $p$ approaches
the total dipole potential
$-(\varrho/2)(\alpha_E E^2 + \alpha_B c^2 B^2)$.
Using the definition (\ref{eq:theta}) of $\Theta^\alpha_\nu$
and the quasi-conservation law (\ref{eq:cl}) 
we obtain
\begin{equation}
\partial_\alpha T_{\mathrm{Mk}\nu}^\alpha =
-\partial_\alpha \sqrt{-g_A}\,\Theta^\alpha_\nu
+\partial_\nu p\,.
\end{equation}
This genuine conservation law suggests that the total 
energy-momentum tensor $T^\alpha_\nu$
of light and matter is
\begin{equation}
\label{eq:total}
T^\alpha_\nu = 
 \sqrt{-g_A}\,\Theta^\alpha_\nu +
T_{\mathrm{Mk}\nu}^\alpha -p\,\delta_\nu^\alpha 
\end{equation}
that contains the Minkowski tensor 
(\ref{eq:mk}) as a major building block.
We use Eqs.\
(\ref{eq:wa}), (\ref{eq:udef}), (\ref{eq:rho}),  (\ref{eq:theta})
and (\ref{eq:pp}) to arrive at the expression
\begin{equation}
T^\alpha_\nu = 
\varrho\,\hbar (\partial_\nu S) c u^\alpha
+ H^{\alpha\alpha'} F_{\alpha'\nu}
+ \frac{\varepsilon_0}{4}
F_{\alpha'\beta'}F^{\alpha'\beta'} \delta_\nu^\alpha \,.
\end{equation}
The components $T^0_l$ with $l\in\{1,2,3\}$ constitute the 
momentum density ${\bf g}$. 
We obtain 
\begin{equation}
{\bf g} = \varrho\,u^0\hbar\nabla S + {\bf D}\times{\bf B} \,,
\end{equation}
which agrees with Eq.\ (\ref{eq:g1}), 
apart from the factor of 
$u^0 = (1-u^2/c^2)^{-1/2}$
that approaches unity in the non-relativistic limit.

\section{Abraham tensor.}

Formulae (\ref{eq:g1}-\ref{eq:u})
illustrate the ambiguity of the 
electromagnetic momentum in media
and the connection 
between the Abraham and Minkowski momenta.
We expect that the fully relativistic 
Abraham and Minkowski tensors
are connected in a similar way.
In order to show this, 
consider the contravariant tensor 
\begin{eqnarray}
\sqrt{-g_A}\Theta^{\alpha\nu}
&=& 
\sqrt{-g_A}\Theta^\alpha_{\nu'} g^{\nu\nu'}
\nonumber\\
&=&
\sqrt{-g_A}\, mc^2 |\psi|^2 w^2 u^\alpha u^\beta g_{\beta\nu'}^A
g^{\nu\nu'}
\nonumber\\
&=&
mc^2\varrho w g^A_{\beta\nu'} g^{\nu\nu'} u^\alpha u^\beta
\,.
\end{eqnarray}
According to definition (\ref{eq:atommetric}) of the atomic metric,
$mc^2\varrho w g^A_{\beta\nu'}$ depends on the free-field
Lagrangian and the free energy-momentum tensor 
(\ref{eq:tensor}) with
the coefficients $a$ and $b$ that we express in terms 
(\ref{eq:ab}) of the 
permittivity $\varepsilon$ and the permeability $\mu$.
In addition, we use 
\begin{eqnarray}
\lefteqn{\left(2-\frac{2}{\mu}\right){\mathscr L}_F}
\nonumber\\
&=&
2p-\frac{2}{\mu}\,{\mathscr L}_F  -
\frac{1}{2}\,F_{\alpha\beta} H^{\alpha\beta}
\nonumber\\
&=&
2p-\left(\varepsilon -\frac{1}{\mu}\right)
\varepsilon_0 F_{\alpha\alpha'}F_{\beta\beta'}
u^{\alpha'} u^{\beta'} g^{\alpha\beta} \,,
\end{eqnarray}
and obtain the result
\begin{equation}
\sqrt{-g_A}\Theta^{\alpha\nu}
= 
[(mc^2 w\varrho +2p) \,u^\nu -(\varepsilon\mu-1)\Omega^\nu]
u^\alpha
\,.
\end{equation}
Here $\Omega^\nu$ abbreviates
\begin{eqnarray}
\Omega^\nu
&=& 
\frac{\varepsilon_0}{\mu} \left(
u^\nu F_{\alpha\alpha'} F_{\beta\beta'} u^{\alpha'} u^{\beta'} 
g^{\alpha\beta} + F_{\beta'\beta}F^{\nu\beta'} u^\beta
\right)
\nonumber\\
&=&
\frac{\varepsilon_0}{\mu}\, F_{\alpha\alpha'}u^{\alpha'}
\left(
u^\nu g^{\alpha\gamma}g^{\beta\beta'} F_{\gamma\beta'} u_\beta
+ g^{\nu\nu'}g^{\alpha\gamma} F_{\nu'\gamma}
\right)
\nonumber\\
&=&
\frac{\varepsilon_0}{\mu}\, F_{\alpha\alpha'}u^{\alpha'}
\left(
u^\nu g_F^{\alpha\gamma}g_F^{\beta\beta'} F_{\gamma\beta'} 
u_\beta
+ g_F^{\nu\nu'}g_F^{\alpha\gamma} F_{\nu'\gamma}
\right)
\nonumber\\
&=&
F_{\alpha\alpha'}u^{\alpha'}
\left(u^\nu H^{\alpha\beta} u_\beta + H^{\nu\alpha}\right)
\nonumber\\
&=&
F_{\alpha\alpha'}u^{\alpha'}u_\beta
\left(H^{\alpha\beta} u^\nu + H^{\beta\nu} u^\alpha +
H^{\nu\alpha} u^\beta \right) 
\end{eqnarray}
that is known as Abraham's {\it Ruhstrahl} \cite{Abraham}.
Finally, 
we express the pressure (\ref{eq:p}) in terms of the norm $w$
using the Hamilton-Jacobi equation (\ref{eq:hj}),
\begin{equation}
p= \frac{mc^2\varrho}{2}\,\left(\frac{1}{w}-w\right)
\,,
\end{equation}
invert this relationship,
\begin{equation}
mc^2w\varrho = \sqrt{m^2c^4\varrho^2 + p^2} - p
\approx mc^2\varrho -p \,,
\end{equation}
and arrive at the representation
\begin{equation}
\sqrt{-g_A} \Theta^{\alpha\nu} = 
\left[(mc^2\varrho+p)u^\nu - (\varepsilon\mu-1)\Omega^\nu\right] 
u^\alpha \,.
\end{equation}
In this way we obtain for the 
total energy-momentum tensor (\ref{eq:total}) the expression
\begin{equation}
T^{\alpha\beta} = T_{\mathrm{Ab}}^{\alpha\beta} + 
(mc^2\varrho +p)u^\alpha u^\beta - p g^{\alpha\beta}
\end{equation}
that describes the energy-momentum balance
between a fluid \cite{LL6} 
under the dielectric pressure $p$
and the electromagnetic field 
in terms of the Abraham tensor
\cite{Abraham,Leo}
\begin{equation}
\label{eq:abr}
T_{\mathrm{Ab}}^{\alpha\beta} = 
T_{\mathrm{Mk}}^{\alpha\beta} - 
(\varepsilon\mu-1) u^\alpha \Omega^\beta \,.
\end{equation}

\section{Conclusions.}
 
We derived two different forms of the energy-momentum 
balance in quantum dielectrics:
the total energy-momentum tensor can be divided into
the Minkowski tensor (\ref{eq:mk}) and the
pseudo-energy-momentum tensor $\Theta_\nu^\alpha$
of the dielectric, apart from a pressure term, or,
alternatively,
into the Abraham tensor (\ref{eq:abr}) 
and the energy-momentum of a fluid
under the dielectric pressure (\ref{eq:p}).
The momentum components of $\Theta_\nu^\alpha$
depend on the phase of the quantum dielectric,
whereas the velocity ${\bf u}$ enters the fluid-mechanical
component of the total energy-momentum tensor.
Due to the R\"ontgen interaction in dielectric media
\cite{Roentgen}
the momentum derived from the gradient of the phase
differs from the mechanical momentum $m{\bf u}$,
a difference that is usually small and often negligible,
whereas the Abraham and Minkowski momenta
differ by the refractive index squared.
We thus arrive at the remarkable conclusion
that the Abraham-Minkowski controversy \cite{Debate}
has its root in the R\"ontgen interaction \cite{Roentgen}.

I thank 
Stephen Barnett,
Iwo Bialynicki-Birula,
Julian Henn,
Rodney Loudon,
Ralf Sch\"utzhold,
Thomas Seligman,
Ilya Vadeiko,
and
Valeri Yakovlev
for illuminating conversations 
on the momentum of light in media.
This paper was supported by the Leverhulme Trust,
the Alexander von Humboldt Foundation and the
Centro Internacional de Ciencias in Cuernavaca.


\begin{thebibliography}{99}
\bibitem{Flaw}
The calculation of the energy-momentum tensor 
in Ref.\ \cite{Leo}
relied on the variation of the total 
Lagrangian of light and dielectric matter
with respect to the background metric.
As an intermediate step, 
a result by Gordon \cite{Gordon}
for media with fixed dielectric properties
was wrongly applied.
\end{thebibliography}
\end{document}